\documentclass{article}
\usepackage{arxiv}

\usepackage[utf8]{inputenc} 
\usepackage[T1]{fontenc}    
\usepackage{hyperref}       
\usepackage{url}            
\usepackage{booktabs}       
\usepackage{amsmath}        
\usepackage{amsfonts}       
\usepackage{nicefrac}       
\usepackage{microtype}      
\usepackage{graphicx}
\usepackage{multirow}       
\usepackage[numbers,sort&compress]{natbib}
\setcitestyle{square,comma}
\usepackage{doi}

\newcommand{\mr}[1]{\mathrm{#1}}

\title{Preliminary Exploration of Large Models Based on Self-Supervised Learning for Health Monitoring of Power Electronic Converters}

\date{} 					

\author{Xinyuan Liao\\
	Department of Electrical and Electronic Engineering\\
	The Hong Kong Polytechnic University\\
	Hong Kong, China \\
	\texttt{xin-yuan.liao@connect.polyu.hk} \\
	\And
	Xiangtian Deng \\
	School of Automation\\
	Wuhan University of Technology\\
	Wuhan 430072, China \\
	\texttt{dengxt@whut.edu.cn} \\
    \And
	Yang Liu \\
	School of Artificial Intelligence and Automation\\
	Huazhong University of Science and Technology\\
	Wuhan 430070, China \\
	\texttt{yangliu30@hust.edu.cn} \\
    \And
	Zhaobing Cai \\
	Three Gorges Intelligent Control Technology Co., Ltd.\\
	Wuhan 430072, China \\
	\texttt{cai\_zhaobing@ctg.com.cn} \\
    \And
	Haoran Wang \\
	Three Gorges Intelligent Control Technology Co., Ltd.\\
	Wuhan 430070, China \\
	\texttt{wang\_haoran@ctg.com.cn} \\
    \And
	Yi Zhang \\
	Department of Electrical and Electronic Engineering\\
	The Hong Kong Polytechnic University\\
	Hong Kong, China \\
	\texttt{yiz@ieee.org} \\
}

\renewcommand{\headeright}{}
\renewcommand{\undertitle}{}
\renewcommand{\shorttitle}{}

\hypersetup{
pdftitle={A template for the arxiv style},
pdfsubject={q-bio.NC, q-bio.QM},
pdfauthor={David S.~Hippocampus, Elias D.~Striatum},
pdfkeywords={First keyword, Second keyword, More},
}

\begin{document}
\maketitle

\begingroup
\renewcommand{\thefootnote}{}
\footnotetext{Accepted by Proceedings of the 2nd Electrical Artificial Intelligence Conference, EAIC 2025}
\addtocounter{footnote}{-1}
\endgroup

\begin{abstract}
	Current artificial intelligence (AI) methods for power electronics converters make limited use of the abundant unlabeled data generated during real-world operation. This underexplored resource holds significant potential for advancing model performance. To address this gap, this paper presents PE-TSFM, a time-series foundation model tailored for power module health monitoring. The model adopts the PatchTST architecture combined with self-supervised learning and is trained on more than 143 million timestamps, where over 98\% of this training data is unlabeled and collected directly from an operating traction converter. In degradation classification tasks, PE-TSFM achieves an accuracy of 87\%, considerably surpassing five conventional time-series AI models whose performance ranges from 27\% to 45\%. These results demonstrate the model's strong few-shot learning capability and generalization across different operating conditions. This work promotes a shift from narrowly focused models in power electronics toward unified, scalable foundation models that can unlock value from large-scale unlabeled operational data.
\end{abstract}


\section{Introduction}\label{sec:intro}

With the large-scale deployment of renewable energy, distributed power grids have attracted extensive attention, in which power electronic converters play a key role~\cite{ref1}. Such grids are geographically sparse and heavily converter-dependent, so the centralized manual operation-and-maintenance strategy adopted by the traditional power generation industry becomes difficult to apply under these conditions. How to provide unified and continuous condition monitoring for a large number of converters is therefore an urgent problem~\cite{ref2,ref3,ref4}. Artificial intelligence (AI) offers a promising direction for the condition monitoring of power electronic converters~\cite{ref5}. However, the demand for large amounts of aging-related labeled data remains the main obstacle to their wide industrial adoption. For example, Ref.~\cite{ref6} collected 10{,}000 labeled samples for open-circuit fault detection, whereas Ref.~\cite{ref7} generated 14{,}000 simulation-based samples under 20 different operating conditions. In practice, aging-related labeled data are difficult to obtain. Moreover, the performance of such methods often degrades substantially under unseen operating conditions. This is a serious concern for industry, because collecting data covering all possible operating conditions and manually labeling them is practically infeasible.

This long-standing limitation raises the core question addressed in this paper: can a unified representation that generalizes across operating conditions be learned from massive unlabeled time-series data? If solved effectively, the dependence on labeled data could be considerably reduced and the value of unlabeled operational data could be released, leading to more practical and transferable industrial condition-monitoring models.

\begin{figure*}
    \centering
    \includegraphics[width=165mm]{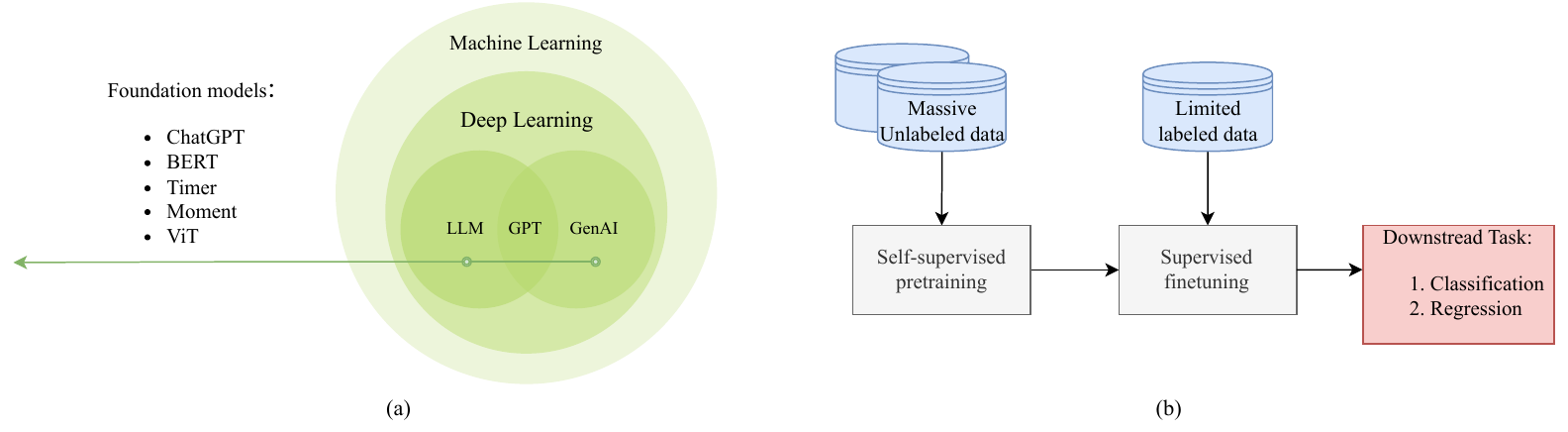}
    \caption{(a) The relationship between the foundation model and traditional AI; (b) the pretraining-finetuning pipeline of the foundation model.}
    \label{fig:concept}
\end{figure*}

Inspired by recent breakthroughs such as BERT~\cite{ref8} and ChatGPT~\cite{ref9}, the AI community has been gradually shifting from task-specific models toward foundation models. As illustrated in Fig.~\ref{fig:concept}, a foundation model~\cite{ref10} is a deep learning model pre-trained by self-supervised learning on massive unlabeled data, which can then be adapted to various downstream tasks with a small amount of labeled data (i.e., fine-tuning). Although foundation models have achieved remarkable performance in natural language processing, time-series analysis, and computer vision, their potential in power electronics remains largely unexplored.

Motivated by this, this paper proposes PE-TSFM, a time-series
foundation model for power module health monitoring. The main
contributions are threefold:
\begin{enumerate}
  \item A foundation-model framework for power electronic systems is
        proposed, which exploits the previously untapped massive
        unlabeled operational data through self-supervised
        pre-training.
  \item A large-scale dataset built on a traction converter is
        constructed, containing 143 million timestamps of
        noninvasive measurements, of which more than 98\% are
        unlabeled. This setting faithfully reflects the scarcity of
        aging labels in industrial practice.
  \item An unseen operating condition is designed to evaluate the
        generalization ability. Five conventional time-series AI
        models achieve only 27\%--45\% accuracy, whereas PE-TSFM
        reaches 87\%.
\end{enumerate}

\section{Practical Challenges of Data-Driven Power Module Health
Monitoring and Dataset Construction}\label{sec:dataset}

\begin{figure*}
    \centering
    \includegraphics[width=165mm]{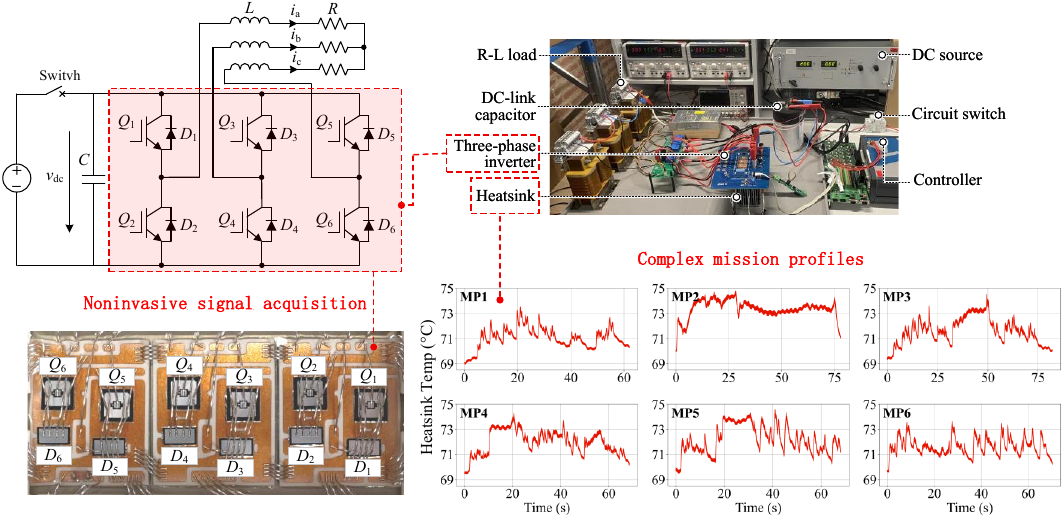}
    \caption{Experimental platform of the traction three-phase inverter.}
    \label{fig:platform}
\end{figure*}

As shown in Fig.~\ref{fig:platform}, the power modules in a traction converter operate under highly dynamic mission profiles, involving frequency variations, thermal cycling, and load fluctuations. These stresses accelerate the aging mechanisms and ultimately lead to electrical or thermal failure. Owing to the safety-critical functions of power modules in electric vehicle and rail transit systems, continuous health monitoring is essential for ensuring system reliability and enabling predictive maintenance.

In industry, the health monitoring of power modules in traction
converters has long faced the following three challenges:
\begin{enumerate}
  \item \emph{Noninvasive monitoring requirement}: traction
        converters in electric vehicle and rail transit systems
        generally do not allow modifications to the existing hardware,
        so all health monitoring must rely on the signals observable
        by the controller, the built-in sensors, or the external ports;
  \item \emph{Massive unlabeled data}: power electronic converters
        continuously generate huge amounts of data during operation,
        whereas labeling the corresponding aging states is extremely
        difficult and costly;
  \item \emph{Unseen operating conditions}: traction converters face
        highly diverse mission profiles, making it infeasible to
        collect training data covering all possible operating
        conditions; models with strong generalization capability are
        therefore urgently required.
\end{enumerate}

To address the above challenges, we construct a dataset based entirely on noninvasive measurements, collected from a 5-kW traction converter, as shown in Fig.~\ref{fig:platform}. All signals are acquired through the existing controller interfaces, built-in sensors, and external ports, without any hardware modification. The dataset contains the controller modulation signal $V_{\mr{m}}$, the DC-bus voltage $V_{\mr{dc}}$, the heatsink temperature $T_{\mr{h}}$, the NTC temperature $T_{\mr{NTC}}$, the three-phase currents $I_{a/b/c}$, and the line voltages $V_{ab}$ and $V_{bc}$, all sampled at 10~kHz. The complete test platform and parameter configuration are detailed in~\cite{ref11}.

To emulate the realistic aging process, the IGBT modules are aged to four levels by standard power-cycling tests. The corresponding on-state voltages are Level~0 (fresh device, 2.907~V), Level~1 (2.969~V), Level~2 (3.025~V), and Level~3 (3.121~V), reflecting progressively increasing degradation. Notably, even at the most severe degradation level (Level~3), the on-state voltage increases by only 0.2~V compared with the fresh device. Such a subtle change is easily masked by the wide dynamic range and fluctuation of the converter operating voltages, making degradation detection under noninvasive measurements particularly challenging.

\begin{table}[!t]
\centering
\caption{Dataset configuration: pre-training on massive unlabeled data
and testing under an unseen condition.}
\label{tab:dataset}
\small
\begin{tabular}{lcccc}
\toprule
Dataset & Label & Task configuration & Total timesteps & Proportion\\
\midrule
\multirow{2}{*}{Training} & Unlabeled & Five mission profiles\,$^{1)}$ & 141\,M & 98.2\%\\
                          & Labeled   & --                                  & 2.56\,M & 1.8\%\\
\midrule
Test & Unlabeled & HWFET (unseen)\,$^{2)}$ & 30\,M & --\\
\bottomrule
\multicolumn{5}{p{0.92\textwidth}}{\footnotesize
$^{1)}$~The five profiles are constructed from five segments of the
U.S. EPA driving cycles, including NYCC (New York City Cycle), UDDS
(Urban Dynamometer Driving Schedule), and LA92.
\quad $^{2)}$~The unseen condition is based on HWFET (Highway Fuel
Economy Test).}
\end{tabular}
\end{table}

Although four explicit aging levels are established through the power-cycling tests, most of the data collected during converter operation remain unlabeled, as summarized in Table~\ref{tab:dataset}. Specifically, power modules at different aging levels are randomly installed on the traction converter platform and operated, generating more than 143 million timestamps. Only 2.56 million timestamps (about 1.8\%) are manually annotated with aging levels, while the remaining 141 million (about 98.2\%) are unlabeled and constitute the unlabeled dataset. This faithfully reflects a common problem in industrial scenarios: operational data are abundant, whereas aging-state labels are extremely scarce.

To better reflect the industrial reality that training data can hardly cover all possible operating conditions, we additionally collect a test dataset whose mission profile never appears during training. As shown in Table~\ref{tab:dataset}, this unseen mission profile differs significantly from the five profiles used for training and is specifically used to evaluate the cross-condition generalization ability.

Conventional AI methods rely heavily on labeled data, whereas in practical industry the labeled data are scarce while the model must generalize to unseen conditions. These constraints pose severe challenges to standard supervised learning. For example, we implement five representative AI methods and test their performance on the collected dataset. As shown in Table~\ref{tab:comparison}, the classification accuracy of these models is only 27\%--45\%, which further highlights the difficulty of achieving reliable generalization under limited supervision.

In summary, the dataset constructed in this section is designed to reflect the practical constraints of power electronics health monitoring: 1) all data must be acquired through noninvasive measurements; 2) aging-related labeled data are extremely limited; and 3) the model must generalize well to unseen operating conditions. Under these constraints, conventional AI methods perform poorly, indicating an urgent need for a new framework that can more effectively exploit the potential of unlabeled data.

\section{The Proposed PE-TSFM Framework}\label{sec:framework}

Based on the dataset constructed in the previous section, this section presents the core contribution of this work: the PE-TSFM framework, which adopts a patch-based Transformer architecture together with a self-supervised pre-training strategy. This framework enables the model to fully exploit massive unlabeled time-series data and, with only a few labeled samples, to achieve excellent generalization to unseen operating conditions.

\begin{figure*}
    \centering
    \includegraphics[width=165mm]{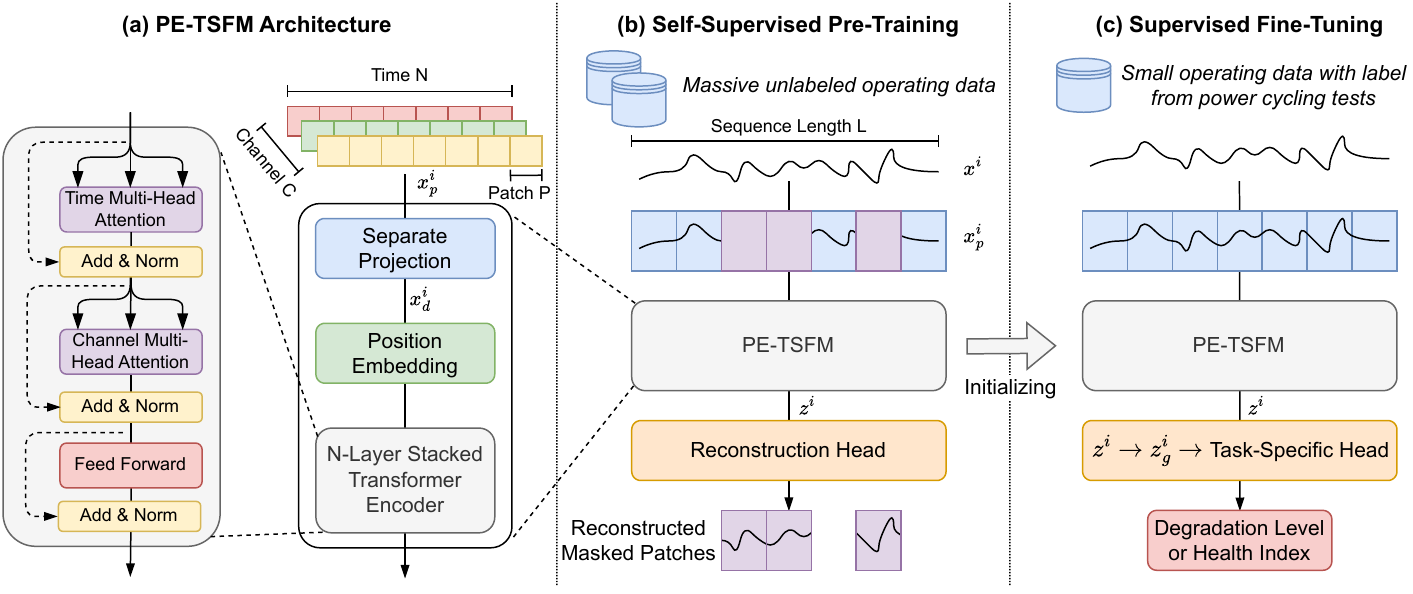}
    \caption{PE-TSFM architecture with the self-supervised pre-training and supervised fine-tuning pipelines for power module degradation monitoring.}
    \label{fig:framework}
\end{figure*}

\subsection{PE-TSFM Structure}\label{sec:structure}

The backbone and overall pipeline of the proposed PE-TSFM are shown in Fig.~\ref{fig:framework}(a). The architecture is inherited from PatchTST~\cite{ref12}, a Transformer model designed for time-series tasks. In this study, the architecture is adapted and enhanced for degradation monitoring of traction converters, so as to better meet the health-monitoring requirements of complex industrial scenarios.

A key contribution of PE-TSFM is the introduction of the patching
mechanism, which converts the raw multivariate time-series data into a
structured sequence of local patches. This operation not only
compresses the sequence length effectively but also preserves
meaningful local temporal dependencies. This design is particularly
effective in power electronics applications, where high sampling rates
often introduce a large amount of noisy or redundant measurements. For
example, the input multichannel time series can be represented as
$x^i\in\mathbb{R}^{C\times L}$, where $C$ denotes the number of
channels (i.e., the number of sensed signals) and $L$ is the sequence
length. The series is reconstructed into a sequence of patches
$x_p^i\in\mathbb{R}^{C\times P\times N}$, where $P$ is the length of
each patch and $N$ is the number of patches. Each patch sequence is
then mapped by a linear projection layer into a higher-dimensional
embedding space, yielding $x_d^i\in\mathbb{R}^{C\times d\times N}$,
where $d$ is the hidden dimension of the model. This process helps
improve the noise immunity of the model, compresses redundant temporal
patterns, and prepares the input for the subsequent attention-based
encoding.

Another key component of PE-TSFM is the temporal--channel dual
attention mechanism with sinusoidal positional encoding~\cite{ref13},
which extracts the degradation-related key patterns from complex
multivariate time-series data. The attention modules enable the model
to selectively focus on the most informative temporal and spatial
features of the multichannel input, which is essential for learning
the long-term dependencies of the degradation process in power
electronic systems. Specifically, the attention mechanism is
formalized as
\begin{equation}
\mr{Attention}(Q,K,V)=\mr{softmax}\!\left(\frac{QK^{\mathrm{T}}}{\sqrt{d}}\right)V,
\label{eq:attention}
\end{equation}
where $Q, K, V\in\mathbb{R}^{d\times N}$ are the query, key, and value
vectors obtained from the patch embedding $x_d^i$ through linear
mappings. Multihead attention (MHA) extends this process to multiple
subspaces to capture diverse dependencies, expressed as
\begin{equation}
\begin{aligned}
\mr{MHA}(Q,K,V)&=\mr{Concat}(\mr{head}_1,\cdots,\mr{head}_n)\,W^{O},\\
\mr{head}_i&=\mr{Attention}(Q_i,K_i,V_i),
\end{aligned}
\label{eq:mha}
\end{equation}
where $Q_i, K_i, V_i\in\mathbb{R}^{(d/n)\times N}$, and $n$ is the
number of attention heads. Since the attention mechanism treats every
time step equally---unlike conventional sequential models such as
recurrent neural networks, which naturally preserve the order of the
time steps---positional encoding is further required so that the model
can distinguish the temporal order. Specifically, the static sinusoidal
positional encoding is adopted:
\begin{equation}
\begin{aligned}
\mr{PE}(pos,2i)&=\sin\!\left(pos/10000^{2i/d}\right),\\
\mr{PE}(pos,2i+1)&=\cos\!\left(pos/10000^{2i/d}\right),
\end{aligned}
\label{eq:pe}
\end{equation}
where $pos\in(0,N\!-\!1)$ denotes the position of each patch in the
sequence, and $i\in(0,d\!-\!1)$ refers to the dimension index of the
high-dimensional embedding $x_d^i$. The model becomes aware of the
temporal order by adding the positional encoding to the original
sequence.

In PE-TSFM, two types of multihead attention are used: 1) the
\emph{temporal MHA}, which captures the temporal dependencies within
each sensor channel; and 2) the \emph{channel MHA}, which models the
relationships across different sensor channels. By stacking multiple
encoder layers, the model finally generates a latent representation
$z^i\in\mathbb{R}^{C\times d\times N}$ of the raw time-series input,
which simultaneously encodes the dynamics along both the temporal and
the sensor dimensions, thereby significantly improving the
generalization ability to unseen operating conditions.

In summary, the PE-TSFM architecture achieves innovations in two
aspects: 1) the \emph{patching mechanism}, which compresses the
redundant time-series data while preserving local temporal dependence;
and 2) the \emph{dual attention mechanism}, which combines temporal
attention and channel attention to comprehensively capture the
correlations within and across sequences. These two designs work
synergistically, enabling robust and generalizable representation
learning in complex and varying power electronic operating
environments, and effectively supporting the health status monitoring
task.

\subsection{Self-Supervised Pre-Training and Supervised
Fine-Tuning}\label{sec:pretrain}

The superior generalization ability of PE-TSFM originates from its
self-supervised pre-training on massive unlabeled data. We borrow the
masked reconstruction learning introduced in BERT, which enables
PE-TSFM to learn a latent representation of the sensor data that is
consistent across operating conditions, thereby significantly
improving its adaptability and transferability to various downstream
tasks.

The pre-training pipeline is shown in Fig.~\ref{fig:framework}(b).
First, a subset of patches is masked, with the masked values set to
the constant 0. Next, PE-TSFM processes the masked patch sequence and
generates the latent representation of the whole sequence. The latent
representation is then fed into a linear reconstruction layer, which
attempts to reconstruct the masked patches. Finally, the
reconstruction error is used as the loss to update the parameters of
the whole model.

The overall pre-training process can be formalized as
\begin{equation}
\theta^{*}=\arg\min_{\theta}\frac{1}{|\mathcal{M}|}
\sum_{j\in\mathcal{M}}
\left\|f_{\theta}\!\left(p^{\mr{masked}}\right)_{j}-p_{j}\right\|_{2}^{2},
\label{eq:pretrain}
\end{equation}
where $\theta$ denotes the trainable parameters of the model,
$\mathcal{M}$ is the index set of the masked patches,
$f_{\theta}(p^{\mr{masked}})_j$ is the reconstructed value of the
$j$-th patch, and $p_j$ is the ground truth of the $j$-th patch.

After pre-training, as shown in Fig.~\ref{fig:framework}(c), the model
enters the fine-tuning stage for the downstream degradation-level
classification task. First, global average pooling is applied to the
sequence representation $z^i$ to extract the global feature vector
$z_g^i\in\mathbb{R}^{d}$. Then, the reconstruction head used in
pre-training is replaced by a linear classification head that maps
$z_g^i$ to a specific degradation level, and the classification loss
is used to continue updating all the parameters of the model so as to
adapt it to the downstream task. It is worth noting that the
fine-tuning stage only requires a small amount of labeled data (see
Table~\ref{tab:dataset}).

This ``pre-training $+$ fine-tuning'' two-stage strategy enables the
model to transfer effectively to downstream health-monitoring tasks
under scarce labeled samples, balancing performance and practicality.

\section{Experimental Results}\label{sec:experiments}

\subsection{Performance Comparison with Baseline Models}
\label{sec:baselines}

To validate the proposed PE-TSFM framework, the dataset in
Table~\ref{tab:dataset} is used. To achieve better cross-condition
generalization, instance normalization~\cite{ref14} is applied to
normalize all data. Instance normalization performs z-score
normalization on each sample; unlike conventional dataset-level
normalization, this sample-level scheme normalizes the test data from
different conditions (including the unseen one) to the same scale,
which effectively improves the cross-condition generalization
performance. It can be summarized as
\begin{equation}
\hat{x}_t^{\,i}=\gamma\cdot\left[x_t^{\,i}-E(x^{i})\right] /
\left[\mr{Var}(x^{i})+\epsilon\right]+\beta,
\label{eq:instance}
\end{equation}
where $x^{i}$ is the $i$-th sample of a single-sensor time series,
$\gamma, \beta\in\mathbb{R}$ are learnable mapping parameters, $E$
denotes the sample mean, and $\mr{Var}$ denotes the sample variance.
All subsequent experiments are run with PyTorch on an Ubuntu server
equipped with an AMD EPYC 7742 CPU and an NVIDIA A100-SXM4-40GB GPU.

First, PE-TSFM is pre-trained by masked reconstruction on the
large-scale unlabeled dataset for 10 epochs with a masking ratio of
30\%. A warmup--cosine learning-rate schedule is adopted, in which the
learning rate starts at $10^{-6}$, increases linearly to a peak of
$2\times10^{-4}$, and finally decays cosine-wise to $10^{-7}$. The
model is then fine-tuned in a supervised manner on the degradation
classification task with all parameters updated, using a small fixed
learning rate of $10^{-5}$ and a batch size of 128. Early stopping
(with the training and validation sets split 8:2) is employed to
prevent overfitting, and gradient clipping is used for training
stability. Following PatchTST, the patch size and the stride are both
set to 16. In addition, two model sizes are trained:
1)~PE-TSFM-3M, with 3.2~M parameters, model dimension
$d_{\mr{model}}=256$, and 4 Transformer encoder layers; and
2)~PE-TSFM-25M, with 25.3~M parameters, model dimension
$d_{\mr{model}}=512$, and 8 Transformer encoder layers. Meanwhile, a
control group without pre-training---trained from scratch with random
initialization---is also set up to verify the effect of pre-training
on model performance. All models are evaluated with three different
random seeds to ensure robust results.

The baseline models for comparison are implemented with the Python
library sktime~\cite{ref15} for time-series analysis, using their
default configurations. For fairness, the training settings of all
methods are kept consistent with the fine-tuning stage of PE-TSFM:
each deep-learning-based model is trained for 100 epochs with a fixed
learning rate of $10^{-5}$.

\begin{table}[!t]
\centering
\caption{Performance comparison with existing methods.}
\label{tab:comparison}
\small
\begin{tabular}{llcc}
\toprule
Model & Training setting & Accuracy (\%) & F1 score (\%)\\
\midrule
MLP (2017)          & from scratch & 27.08 & 26.65\\
ResNet (2017)       & from scratch & 39.65 & 39.23\\
LSTM-FCN (2019)     & from scratch & 35.93 & 33.55\\
TapNet (2020)       & from scratch & 35.92 & 35.68\\
MiniRocket (2021)   & from scratch & 44.85 & 42.38\\
\midrule
PE-TSFM-3M          & from scratch & 55.92 & 53.83\\
PE-TSFM-25M         & from scratch & 66.86 & 66.07\\
PE-TSFM-3M          & pre-trained  & 76.97 & 75.72\\
\textbf{PE-TSFM-25M}& \textbf{pre-trained}  & \textbf{87.17} & \textbf{87.13}\\
\bottomrule
\end{tabular}
\end{table}

Table~\ref{tab:comparison} summarizes the average results over three
runs, where PE-TSFM is compared with five representative time-series
classification models: the multilayer perceptron (MLP), the residual
network based on one-dimensional convolutions (ResNet), the long
short-term memory--fully convolutional network (LSTM-FCN), the
time-series attention prototype network (TapNet), and MiniRocket. In
terms of both accuracy and F1 score, the proposed PE-TSFM
significantly outperforms all the baselines. Notably, the pre-trained
PE-TSFM-25M achieves an accuracy of 87.17\% and an F1 score of
87.13\%, far exceeding the best-performing baseline (MiniRocket,
44.85\% accuracy).

These results reveal two main limitations of task-specific models:
1)~the lack of a unified representation leads to poor generalization
across operating conditions; and 2)~high sensitivity to training
initialization results in large performance variance. In contrast,
benefiting from self-supervised pre-training and the architectural
design, PE-TSFM exhibits significant advantages in few-shot learning
and cross-condition generalization.

\subsection{Pre-Training and the Scaling Law}\label{sec:scaling}

To demonstrate the positive effect of self-supervised pre-training on
PE-TSFM, two additional control groups are trained, whose model sizes
are identical to the full PE-TSFM but whose parameters are randomly
initialized without self-supervised pre-training, with all other
training details the same as in Section~\ref{sec:baselines}. As can be
seen from Table~\ref{tab:comparison}, without pre-training, PE-TSFM
achieves an accuracy of only 55\% and 66\% and an F1 score of only
53\% and 66\%, whereas pre-training improves the accuracy and F1 score
by up to 21 and 22 percentage points, respectively. These results
prove that self-supervised pre-training has a significant positive
effect on the generalization performance under unseen operating
conditions.

In addition, we observe that the larger foundation model exhibits
better cross-condition generalization in the tests, which is
consistent with the scaling law observed in large language models. The
scaling law~\cite{ref16} generally refers to the relationship between
model performance and certain key factors, such as model size, dataset
size, and computational budget. This work focuses on the relationship
between model performance and model size. In large language models,
when a model with a limited number of parameters $N$ is trained to
convergence on a sufficiently large dataset, the scaling law can be
empirically defined as
\begin{equation}
L(N)=N_c\cdot N^{-a_N},\quad N_c>0,\; a_N>0,
\label{eq:scalingL}
\end{equation}
where $L$ is the performance metric, such as the mean squared error
(MSE) or the root mean squared error (RMSE), $N_c$ is a constant
determined by the model architecture and the task, and $a_N$ is the
scaling exponent reflecting how fast the performance improves as the
number of parameters increases. For classification tasks, the scaling
law of accuracy can be expressed as
\begin{equation}
A(N)=A_{\max}-N_c\cdot N^{-a_N},
\label{eq:scalingA}
\end{equation}
where $A_{\max}$ is the upper bound of the accuracy.

\begin{figure}[t]
    \centering
    \includegraphics[width=85mm]{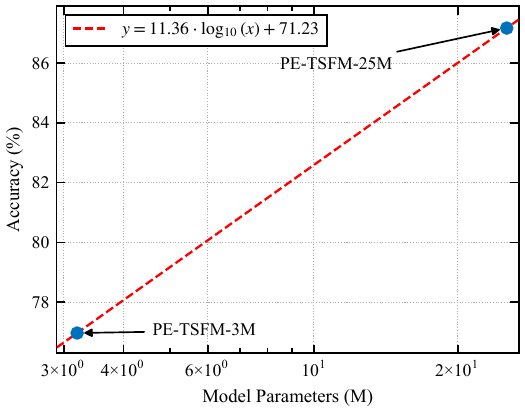}
    \caption{The relationship between the number of model parameters and the accuracy under the same amount of training data (log scale).}
    \label{fig:scaling}
\end{figure}

In this instance, the curve-fitting results indicate that the model
performance follows an approximately logarithmic relationship with the
number of parameters, as shown in Fig.~\ref{fig:scaling}. As the model
scale increases, the performance improves logarithmically, suggesting
that the scaling law also holds for time-series foundation models in
power electronics. Therefore, the model performance can be further
improved by scaling up the number of parameters, indicating the
potential for further expansion.

\section{Conclusion}\label{sec:conclusion}

This paper proposes PE-TSFM, a time-series foundation model for power module health monitoring under realistic converter operating conditions. By fully exploiting massive unlabeled data through self-supervised pre-training, the model learns condition-unified representations that effectively support few-shot learning and generalize excellently to unseen operating conditions. The innovative patching and dual attention designs capture both the temporal and the inter-sensor dependencies. Experimental results show that PE-TSFM achieves a classification accuracy of 87.17\% using only limited labeled data, nearly twice that of the best-performing baseline model. Overall, this work provides a scalable and transferable framework for power electronics health monitoring, which is expected to release the potential value of unlabeled operational data and significantly improve the adaptability of models under complex operating conditions.


\clearpage
\appendix
\section{Award Certificate}\label{app:award}

\begin{figure}[h]
\centering
\includegraphics[width=0.7\textwidth]{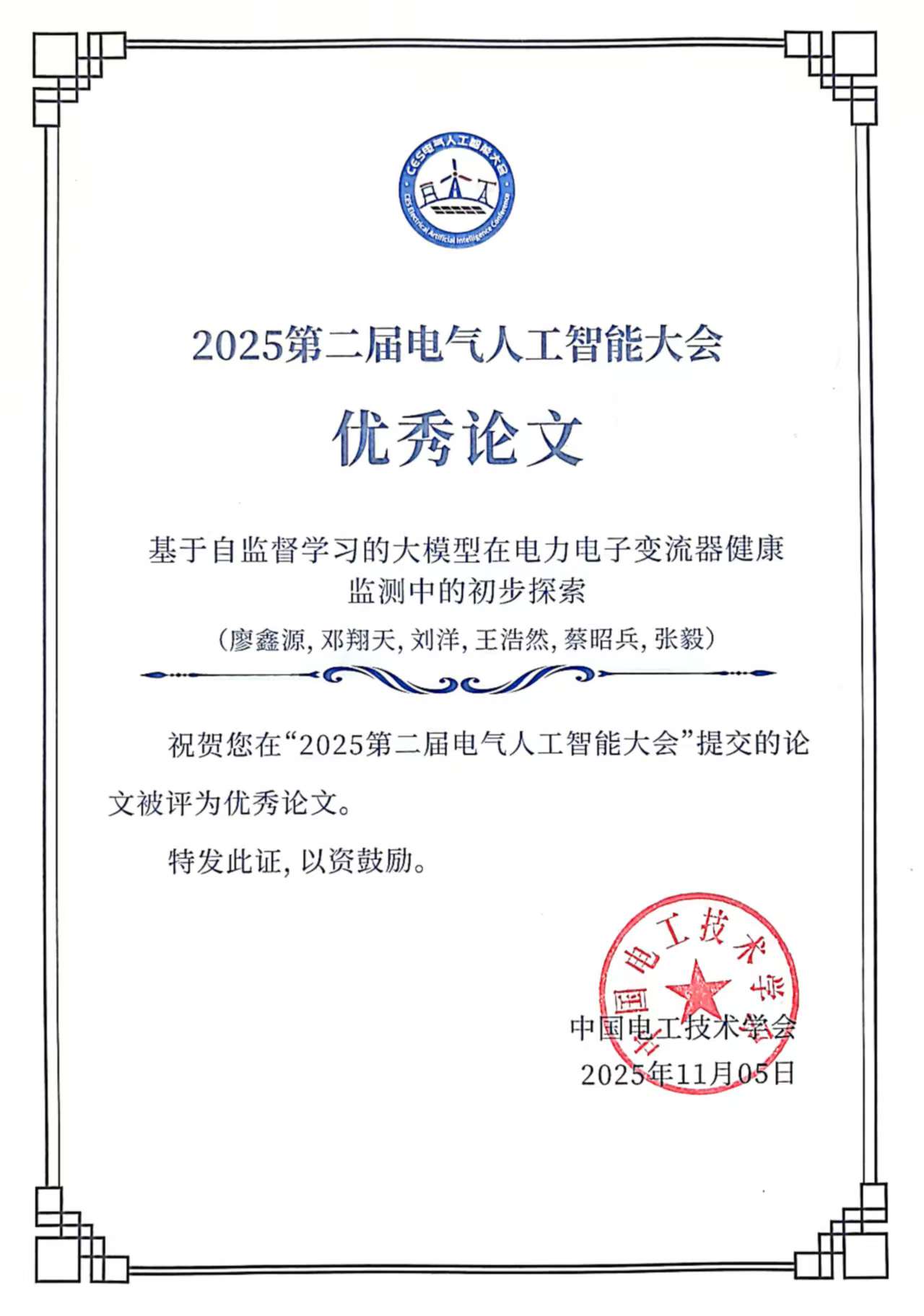}
\caption{Outstanding Paper Award certificate of the 2nd Electrical
Artificial Intelligence Conference (EAIC 2025), issued by the China
Electrotechnical Society on November 5, 2025.}
\label{fig:award}
\end{figure}

This paper received the \emph{Outstanding Paper Award} at the 2nd Electrical Artificial Intelligence Conference (EAIC 2025), conferred by the China Electrotechnical Society on November 5, 2025, as shown in Fig.~\ref{fig:award}.

\begin{thebibliography}{99}
\small

\bibitem{ref1} M. Huang, S. Shu, X. Li, et al., ``Research review on analysis and control of power electronic grid-connected converters for synchronization stability,'' \textit{Transactions of China Electrotechnical Society}, vol. 39, no. 19, pp. 5978--5994, 2024.

\bibitem{ref2} S. Xu, X. Xu, Y. He, et al., ``Open-switch and current sensor fault diagnosis of neutral-point-clamped three-level grid-connected inverter based on adaptive sliding mode observer,'' \textit{Transactions of China Electrotechnical Society}, vol. 39, no. 13, pp. 4066--4078, 2024.

\bibitem{ref3} C. Chen, J. Dai, T. Sun, et al., ``Open-circuit fault diagnosis method of neutral-point-clamped inverter for residual evaluation,'' \textit{Transactions of China Electrotechnical Society}, vol. 39, no. 21, pp. 6877--6895, 2024.

\bibitem{ref4} Q. Wan, J. Yu, Q. Tong, et al., ``IGBT aging failure prediction method based on Bo-BiLSTM network,'' \textit{Electrical Engineering}, vol. 25, no. 3, pp. 1--10, 2024.

\bibitem{ref5} S. Zhao, F. Blaabjerg, and H. Wang, ``An overview of artificial intelligence applications for power electronics,'' \textit{IEEE Transactions on Power Electronics}, vol. 36, no. 4, pp. 4633--4658, 2021.

\bibitem{ref6} Y. An, X. Sun, B. Ren, et al., ``A data-driven method for IGBT open-circuit fault diagnosis for the modular multilevel converter based on a modified Elman neural network,'' \textit{Energy Reports}, vol. 8, pp. 80--88, 2022.

\bibitem{ref7} Y. Liu, A. Sangwongwanich, Y. Zhang, et al., ``A deep learning network based robust fault diagnosis method for IGBT open circuit,'' in \textit{Proc. IEEE 10th International Power Electronics and Motion Control Conference (IPEMC2024-ECCE Asia)}, 2024, pp. 2366--2371.

\bibitem{ref8} J. Devlin, M.-W. Chang, K. Lee, and K. Toutanova, ``BERT: Pre-training of deep bidirectional transformers for language understanding,'' in \textit{Proc. 2019 Conference of the North American Chapter of the Association for Computational Linguistics: Human Language Technologies}, 2019, pp. 4171--4186.

\bibitem{ref9} J. Achiam, S. Adler, S. Agarwal, et al., ``GPT-4 technical report,'' arXiv preprint arXiv:2303.08774, 2023.

\bibitem{ref10} R. Bommasani, D. A. Hudson, E. Adeli, et al., ``On the opportunities and risks of foundation models,'' arXiv preprint arXiv:2108.07258, 2021.

\bibitem{ref11} H. Xia, Y. Zhang, D. Zhou, et al., ``Impact of loss model selection on power semiconductor lifetime prediction in electric vehicles,'' in \textit{Proc. IECON 2022 -- 48th Annual Conference of the IEEE Industrial Electronics Society}, 2022, pp. 1--7.

\bibitem{ref12} Y. Nie, N. H. Nguyen, P. Sinthong, and J. Kalagnanam, ``A time series is worth 64 words: Long-term forecasting with transformers,'' arXiv preprint arXiv:2211.14730, 2022.

\bibitem{ref13} A. Vaswani, N. Shazeer, N. Parmar, et al., ``Attention is all you need,'' in \textit{Advances in Neural Information Processing Systems}, vol. 30, 2017.

\bibitem{ref14} T. Kim, J. Kim, Y. Tae, et al., ``Reversible instance normalization for accurate time-series forecasting against distribution shift,'' in \textit{Proc. International Conference on Learning Representations (ICLR)}, 2021.

\bibitem{ref15} M. L\"oning, A. Bagnall, S. Ganesh, et al., ``sktime: A unified interface for machine learning with time series,'' arXiv preprint arXiv:1909.07872, 2019.

\bibitem{ref16} J. Kaplan, S. McCandlish, T. Henighan, et al., ``Scaling laws for neural language models,'' arXiv preprint arXiv:2001.08361, 2020.

\end{thebibliography}
\end{document}